\documentclass{elsart}

\usepackage{graphicx}
\usepackage{natbib}
\usepackage{amssymb}
\journal{Surface Science}

\usepackage{graphicx}
\usepackage{dcolumn}
\usepackage{bm}

\begin{document}
\begin{frontmatter}

\title{Epitaxy and growth of titanium buffer layers on $Al_2O_3(0001)$}

\author[SVI]{E. S{\o}nderg{\aa}rd\corauthref{cor}}
\corauth[cor]{Corresponding author. }
\ead{elin.sondergard@saint-gobain.com},
\author[SVI]{O. Kerjan},
\author[Saclay]{C. Barreteau} and
\author[GPS]{J.Jupille}
\address[SVI]{Laboratoire CNRS-Saint Gobain, "Surface du Verre et
Interfaces", 93303 Aubervilliers, France.}
\address[Saclay]{DSM/DRECAM/SPCSI, CEA Saclay, F-91 191 Gif sur Yvette, France}
\address[GPS]{Groupe de Physique des Solides, Universit\'{e}s Paris 6 et 7, 140
rue de Lourmel, 75015 Paris, France}

\date{\today}

\begin{abstract}
The structure and growth of thin films of titanium on
$\alpha-\makebox{Al}_2\makebox{O}_3$ at room temperature were
investigated though {\it in situ} RHEED observations. Two
different structures coexists at low coverage. One corresponds to
the $\makebox{Ti}(0001)
\parallel \makebox{Al}_2\makebox{O}_3\makebox{(0001)} $,
$\makebox{Ti} [ 1 \overline{1}00]
\parallel \makebox{Al}_{2}\makebox{O}_{3}[2\overline{1}\overline{1}
0] $ and $\makebox{Ti}[10\overline{1}0]\parallel
\makebox{Al}_2\makebox{O}_3[1\overline{1}00]$ epitaxy of the
$\alpha$ phase of titanium reported before for thick films
prepared at high temperature. The other structure can be explained
by co-existence of $\alpha$ and $\omega$ Ti in thin films.  It was
shown, with the use of tight-binding  total energetic calculations
that, the $\omega$ phase could actually be stabilized by the
substrate. In addition, it was demonstrated that the presence of
this extra structure has a dramatic effect on the epitaxial growth
of the Ag overlayers on the system. This can be the origin of the
non-trivial buffer effect of titanium previously observed.
\end{abstract}

\end{frontmatter}

\section{\label{sec:Intro}Introduction}
Finding surface treatments which can force layer-by-layer growth
is among the key issues  in heteroepitaxy \cite{Bauer}. Indeed,
many attempts have been made to promote the wetting of materials
of different and non interacting nature  such as the metal on
oxide interface \cite{evans,Ti2}. The challenge  is not only to
grow durable interfaces but also to control the morphology of the
metal film as it affects the physical properties of the system.
The subject is of increasing importance as many industrial
applications rely on the mastering of interfaces such as copper
interconnects on oxides for microelectronics and silver on oxides
for optical purposes. As a trend, adhesion energy decreases from
early transition metals to noble metals \cite{campbell,JJ}, due to
the increased  filling of the 3d band \cite{Johnson,Alemany}.
These differences have major consequences regarding the wetting
properties of metals on oxides and lead to a large variety of
growth situations. The thermodynamic condition for two-dimensional
growth of a metal film on a substrate requires the adhesion energy
$\beta$ to be equal to or higher than twice the surface energy of
the metal $\sigma_M$. The way $\beta$ compares to $\sigma_M$ is of
central importance for the growth of the film. While the condition
$ \beta < \sigma_M$ implies $\theta
> 90^\circ$ and a growth in a
three-dimensional fashion up to very high coverage, $\beta >
\sigma_M$ implies $\theta < 90^\circ$ and leads to a quickly
percolating film. Clearly, noble metals and late transition metals
on wide band gap oxides belong to the first category \cite{D.
Chatain}  and early transition metals to the second \cite{Pedden}.
A method often used to improve the wetting of noble metals on
oxide surfaces is pre-deposition of early transition metals
\cite{Köstlmeier}. This so-called buffer effect involves both
electronic and geometric contributions \cite{Köstlmeier}.

\begin{figure}
\includegraphics[width=15cm]{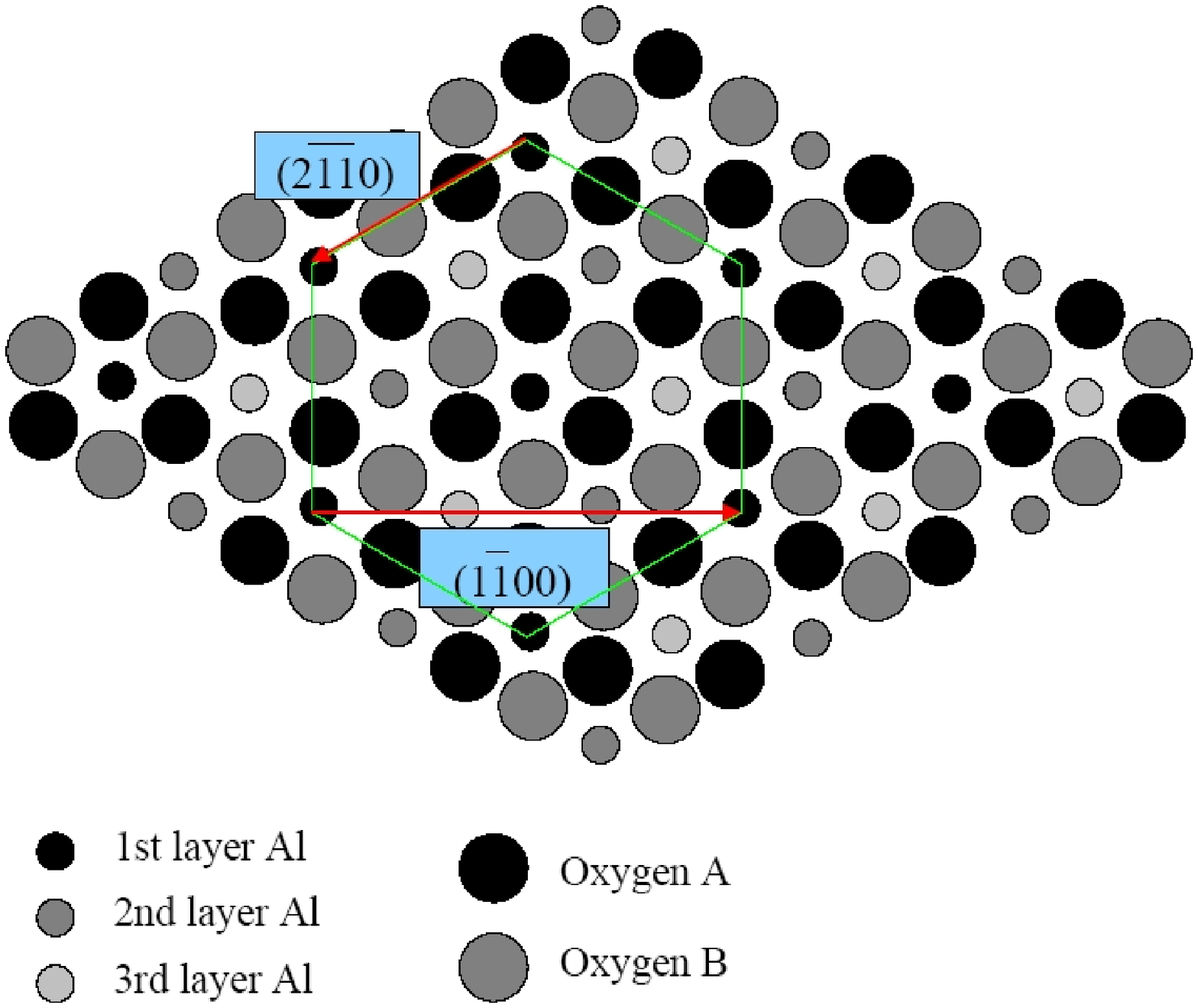}
\caption {Schematic Top view of an Al terminated $\alpha$
Al$_2$O$_3$(0001) surface. The hexagon indicates the position of
the outer Al- layer.} \label{fig:Al2O3}
\end{figure}

In the present work, attention is focused on the effect of
titanium buffers for noble metal growth on alumina, a subject
receiving  attention because of its application in microelectronic
devices \cite{Ti2,Köstlmeier}, and references therein. Suzuki et
al. \cite{Ti1} found by  ion scattering that there is no
interdiffusion at the interface between the titanium overlayer and
the Al-terminated $\alpha -
\makebox{Al}_2\makebox{O}_3\makebox{(0001)}$ substrate. By
performing surface diffraction experiments on a 20 monolayer thick
titanium film annealed to 1100 K, they showed that the epitaxial
relationship was  $\makebox{Ti}(0001)
\parallel \makebox{Al}_2\makebox{O}_3\makebox{(0001)} $,
$\makebox{Ti} [1 \overline{1}00]
\parallel \makebox{Al}_{2}\makebox{O}_{3}[2\overline{1}\overline{1}
0]$ and $\makebox{Ti}[10\overline{1}0]\parallel
\makebox{Al}_2\makebox{O}_3[1\overline{1}00]$. This means that the
triangular lattice of Ti(111)  is rotated by $30^\circ$ with
respect to the triangular lattice formed  by the outer Al atoms of
$\makebox{Al}_2\makebox{O}_3\makebox{(0001)}$, see
figure~\ref{fig:Al2O3}. Dehm et al. \cite{Ti2} came to similar
conclusions through combined transmission electron microscopy
(TEM) and reflection high energy electron diffraction (RHEED)
observations of a 10-100 nm thick titanium film on
$\makebox{Al}_2\makebox{O}_3\makebox{(0001)}$. These authors did
not investigate in details  the crystal structure of the interface
Ti/$\makebox{Al}_2\makebox{O}_3$. Recently, a first principle
calculation of Ti on $\makebox{Al}_2\makebox{O}_3\makebox{(0001)}$
predicted a tendency to intermixing for aluminium and titanium at
the interface  such that the most stable site for the titanium
adatoms is above or in replacement of the surface aluminium atoms
\cite{verdozzi}.  These results have still not been experimentally
verified.

Concerning the buffer effect it was reported in Ref. \cite{Ti2}
that copper grew in a epitaxial manner on thick titanium films
(10-100 nm) on $\makebox{Al}_2\makebox{O}_3\makebox{(0001)}$. The
buffer effect of titanium was indirectly demonstrated on silver
for much thinner films  through optical reflection {\it in situ}
measurements of the silver plasmon \cite{SGR}. It was found that
the use of titanium buffers had a significant influence on the
morphology of silver films, although the effect of the titanium
film was shown to depend highly on its thickness. Very thin
titanium layers ($< 0.3$ nm) hardly produce any change in the
three dimensional growth of silver with respect to the bare
$\makebox{Al}_2\makebox{O}_3$ surface. In contrast,  a titanium
layer of 0.6 nm  leads to a better wetting of silver between
silver and the substrate. However, for higher titanium coverage,
the quality of the wetting decreases, although it remains better
than on the bare surface. Similar results where obtained by
surface diffraction experiments \cite{ISSPICJPD}. The very
intriguing complexity of this scenario can hardly be explained by
the surface energy of titanium alone. In the present work the
structure of the titanium buffer is investigated as a fonction of
its thickness. The corresponding behavior of a silver overlayer is
further discussed.

\section{Experimental}
Single crystalline $\makebox{Al}_2\makebox{O}_3\makebox{(0001)}$
substrates from Ma\-teck Gmbh with a $\leq 0.5^{\circ}$ miscut
were prepared using acetone plus ultrasonic bath followed by a
cleaning in a 10 percent dilution of a standard buffered basic
soap with PH=10.6. After washing in deionized water and drying in
isopropanol vapor, an {\it ex situ} annealing at 1320 K was
performed under an atmospheric pressure of oxygen  to obtain a
good crystallinity. Afterwards, samples were inserted in a
MECA2000 molecular beam epitaxy apparatus, annealed under a
partial pressure of  $5 \cdot 10^{-5}$ torr $\makebox{O}_2$  to
get rid of carbon contamination, first at 820 K over a couple of
hours and then at 1070 K for one hour.  Samples were cooled under
oxygen prior to transfer to the evaporation chamber where the base
pressure was $1 \cdot 10^{-10}$ torr. In a similar vacuum chamber
equipped with a photoemission analyser, is was checked that this
treatment leads to clean
$\makebox{Al}_2\makebox{O}_3\makebox{(0001)}$ surfaces \cite{SGR}.

Titanium and silver films were evaporated on the samples using a
Telemark electron gun. The flux was between 0.08-0.1 nm  per min
and the samples were held at 400 K. As titanium is a very reactive
material special care was taken to keep the chamber pressure
within the $10^{-10}$ torr range during evaporation. The thickness
was measured with a quartz microbalance  calibrated using {\it ex
situ} electron microprobe measurements to obtain the number of
atoms deposited. To eliminate any distorsions of the diffraction
pattern due to stray fields from the electron guns, special care
was taken to align the optical center of the RHEED with the
electron guns in operation condition. The evolution of the growth
was followed by a Staib RHEED instrument and the diffraction
patterns were on a CCD camera. During the deposition any physical
movement of the samples used for the lattice measurements were
avoided to keep the substrate reference for the diffraction
patterns. When analysing short lattice spacings, for which the
diffraction patterns would be outside the screen, the scattered
beams were deflected on the screen by the means of an
electrostatic device. It was verified several times that this
method does not introduce any errors on the positions of the
diffraction peaks on the bare sapphire substrate.

\section{Results}
The following section first presents the results and the
discussion related to the growth of titanium on $\makebox{Al}_2
\makebox{O}_3$. The influence of titanium on the growth of silver
is then briefly considered. Directions along which RHEED analysis
is performed are always given with respect to the crystallographic
axes in  figure ~\ref{fig:Al2O3}. Distances in real and reciprocal
space will be labelled $l$ and $g$ respectively.

\subsection{Growth of the titanium film}
\begin{figure}
\includegraphics[width=19cm]{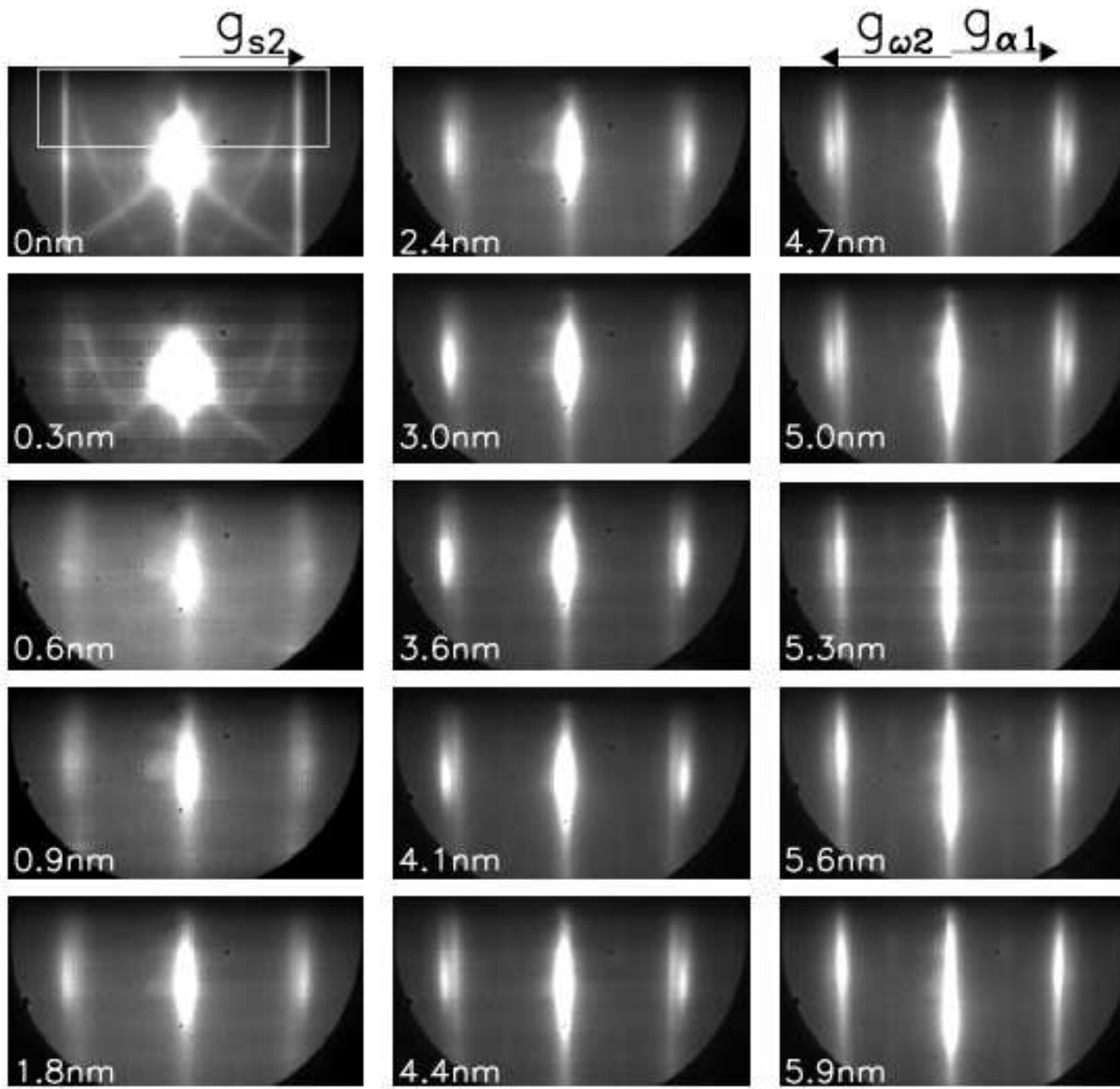}
\caption { RHEED images showing the growth of the Ti film with the
beam along the $Al_2O3[1\overline{1}00]$ axis.} \label{fig:axe1}
\end{figure}

Figure~\ref{fig:axe1} shows the typical evolution of the RHEED
pattern of a titanium layer collected along the
$\makebox{Al}_2\makebox{O}_3 [ \overline{1}100 ]$ zone axis which
means that the observed surface streaks correspond to the
$\makebox{Al}_2\makebox{O}_3 [ 2\overline{11}0 ]$ axis. The film
starts growing with a reciprocal lattice parameter close to the
substrate lattice, $g_{s2}$. Up to 1 nm, the broad and not very
intense diffraction rods indicate a low degree of order. Upon
increasing the titanium coverage,  the diffraction pattern
globally gains intensity. Between approximately 1.8- 6 nm, two
separated lattice parameters $g_{\omega 2}$ and $g_{\alpha 1}$
coexist. At lower coverage, the external rod $g_{\omega 2}$ is the
most intense (figure~\ref{fig:axe1}, 0.3-3.6 nm.). For increasing
coverage the internal rod $g_{\alpha 1 }$ increases in intensity
to finally dominate the pattern (figure~\ref{fig:axe1}, 5.6 nm).
Note that the thickness where the switch in intensity between
$g_{\omega 2}$ and $g_{\alpha 1}$ occurs depends on the sample.
For some crystals it is impossible to get a correct diffraction
pattern below 2 nm of Ti - for others the switching between the
lattice spacings $g_{\omega 2}$ and $g_{\alpha 1}$ is already
observed at a lower coverage. But, in all cases,  in-plane
rotations of $360^{\circ}$ show that the symmetry of the structure
is 6 over the whole thickness range.

\begin{figure}
\includegraphics[width=12cm]{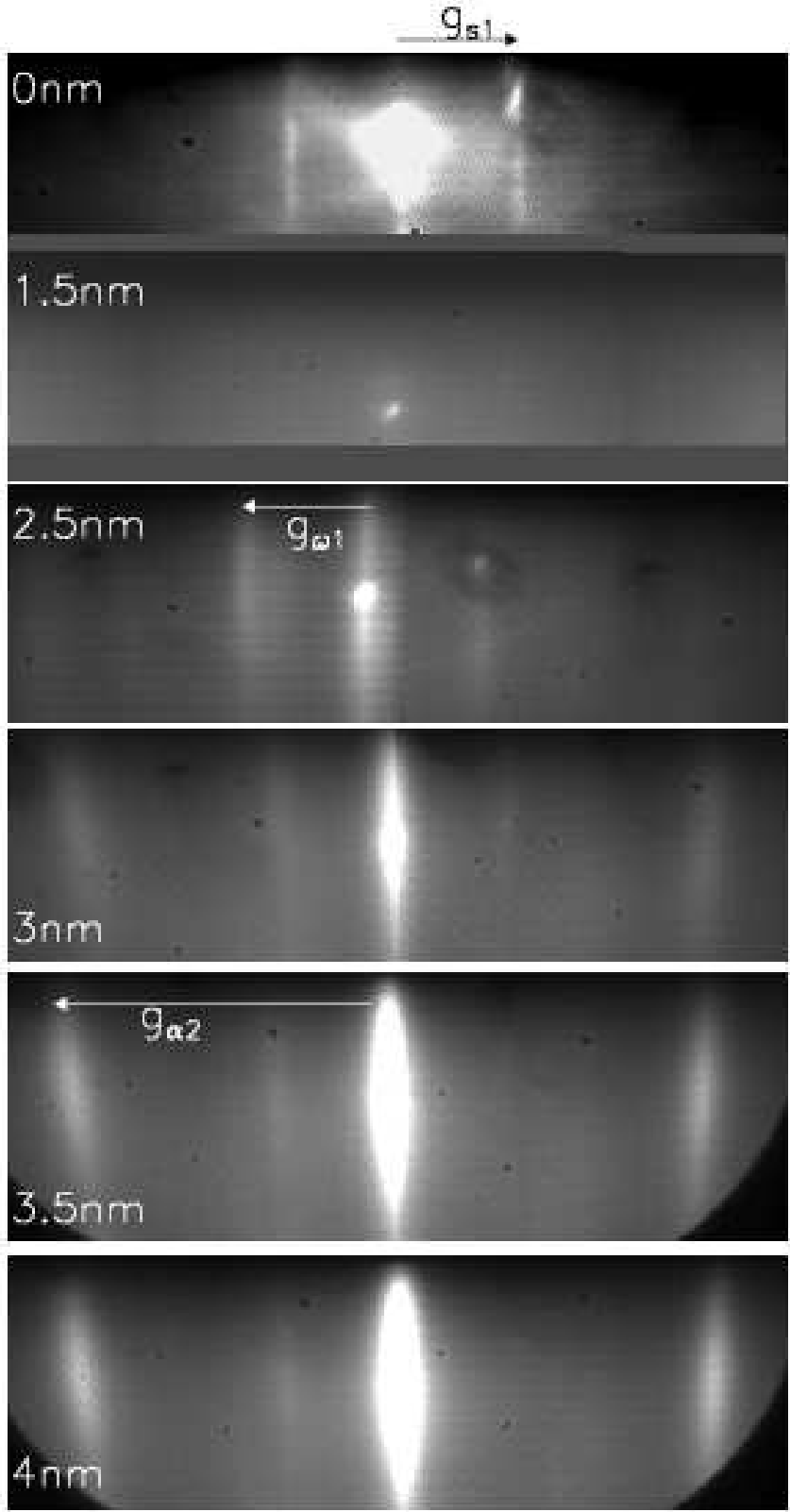}
\caption {RHEED images showing the growth of the Ti film with the
beam along $Al_2O_3[2\overline{11}0]$. The images presented were
obtained as the recombination of two deflected images in order to
observe the diffraction peak due to the $Ti[10\overline{1}0]$.}
\label{fig:axe2}
\end{figure}

To collect the peak corresponding to Ti$[ 10\overline{1}0 ]$, the
RHEED pattern of the growing titanium was also recorded along the
$\makebox{Al}_2\makebox{O}_3 [2\overline{11}0 ]$ azimuth. This was
delicate because the working distance in the RHEED setup is mainly
aimed at analysing large lattice spacings. Small lattice spacings
can only be observed by deflection of the diffraction pattern on
the CCD camera. Therefore, the diffraction patterns were collected
by halfs. To get a complete image the two halfs were recombined
using the (00) streaks as a reference (figure~\ref{fig:axe2}). The
corresponding lattice parameters were obtained as the mean value
found from the half images. The diffraction patterns in the
$\makebox{Al}_2\makebox{O}_3 [ 2\overline{11}0 ]$ zone axis were
in general of low intensity. Interestingly, the patterns obtained
around 2 nm of titanium are close to those of the clean substrate
(figure~\ref{fig:axe2}). This pattern disappears as the streaks
corresponding to the Ti$[ 10\overline{1}0]$ appear. A full
rotation of the sample reveals a six fold symmetry of the pattern.

\begin{figure}
\includegraphics[width=13cm]{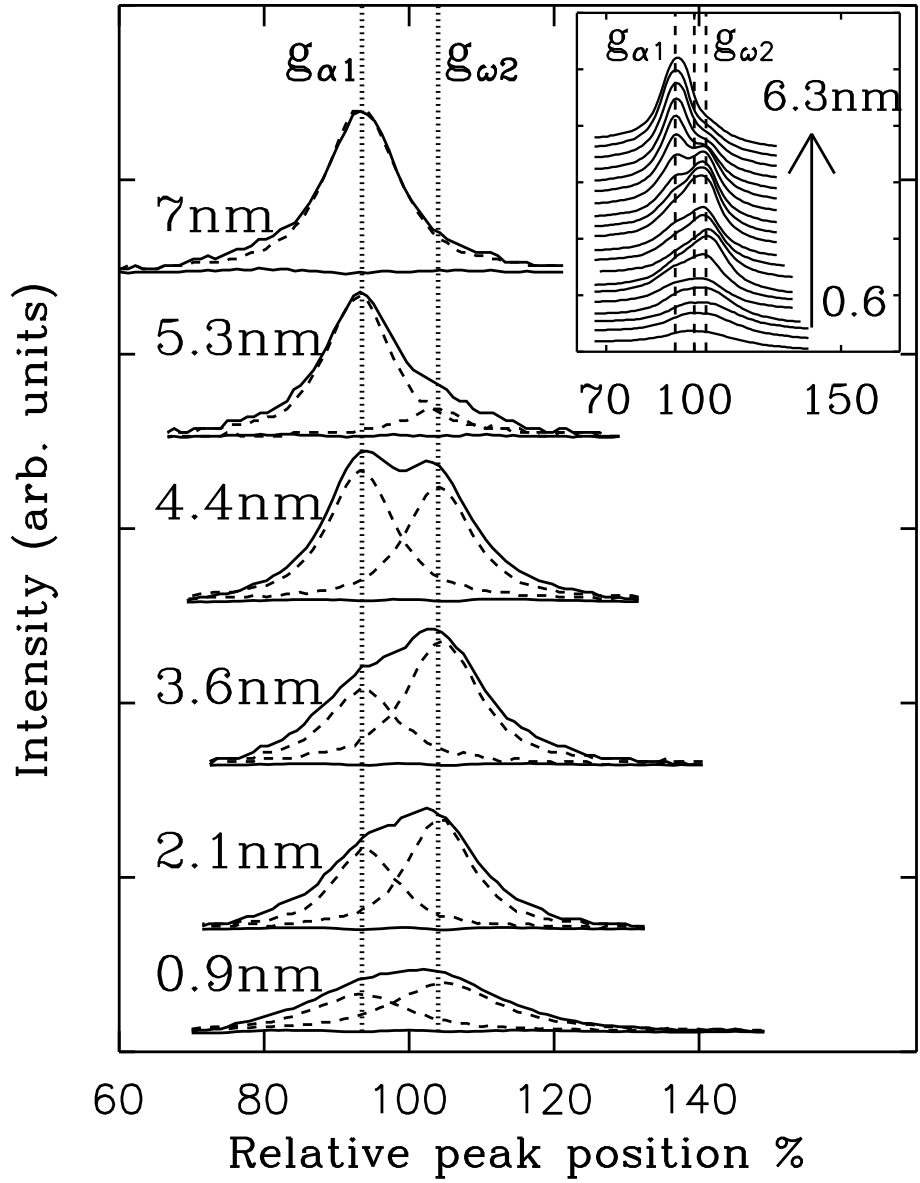}
\caption {The intensity profiles of the diffraction streaks
recorded along $Al_2O_3[1\overline{1}00]$ axis
(figure~\ref{fig:axe1}) for titanium films of thickness ranging
from 0.9 to 6 nm and  their corresponding fits with Voigt
functions. The gradual shift in intensity from $g_{\omega 2}$ to
$g_{\alpha 1}$ positions is seen on the inserted figure.}
\label{fig:peaks}
\end{figure}

\begin{figure}
\includegraphics[width=13cm]{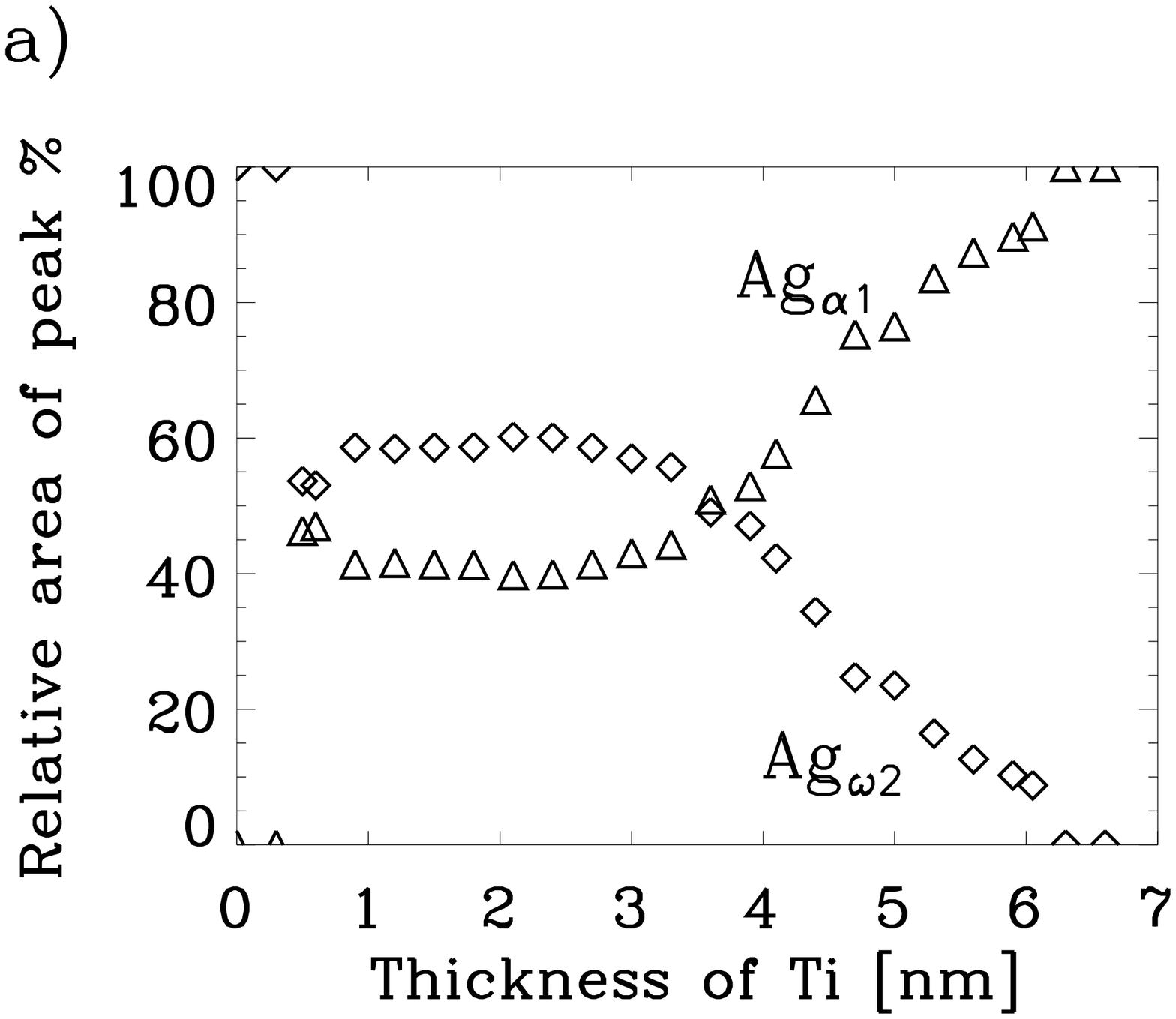}
\includegraphics[width=13cm]{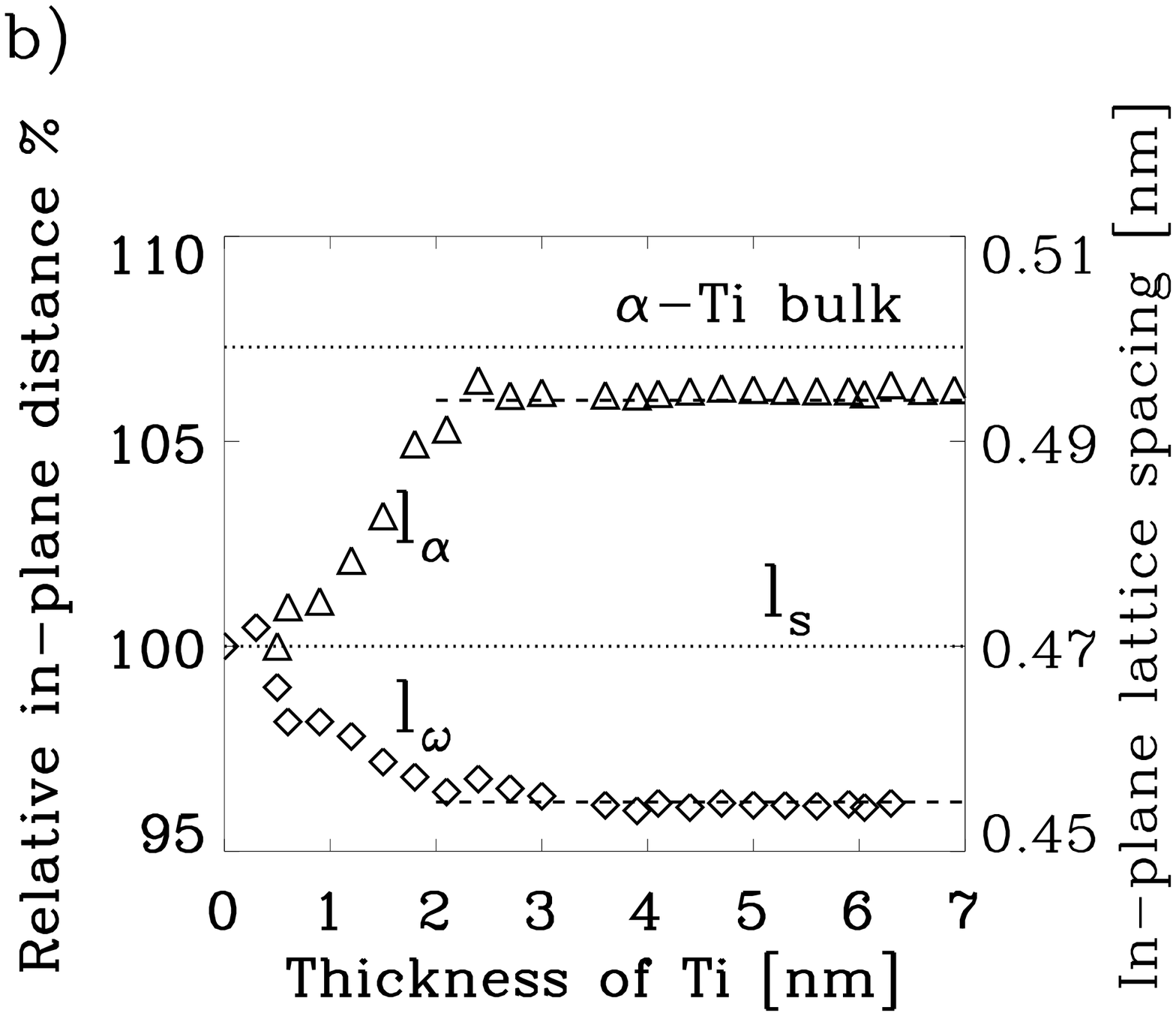}
\caption {a) The area of the diffraction peaks of
figure~\ref{fig:axe1} given by the fitting procedure. b) The
fitted lattice spacings, $l_{\alpha}$ and $l_{\omega}$, relative
to the substrate, $l_s$.} \label{fig:lat}
\end{figure}

The profiles of the diffraction features were extracted to analyse
the pattern in the $\makebox{Al}_2\makebox{O}_3 [ 1 \overline{1}00
]$ zone axis.  In order to smooth the noise of the data, three
photos recorded for each film thickness were summed and
intensities were cumulated along the diffraction peaks. To avoid
zones where the diffraction streaks can be curved only the part of
the diffraction pattern close to the shadow edge of the sample was
used. Intensity profiles are showed in figure ~\ref{fig:peaks}
(upper right corner). The analysis of this data is complicated by
the presence of two overlapping and broad peaks. Nevertheless,
since in a wide coverage range, the streaks are split into two
lines separated by a valley, it has been chosen to decompose the
profiles into two components over the whole titanium coverage
range. Diffractions peaks are Lorenzian functions but deviations
from this are expected when working with a CCD camera. Therefore
several types of functions were tested to obtain the best possible
model over as large a range of coverage as possible. It proved to
be a Voigt function with a 80\% ratio of the Lorenzian to
Gaussian. All the diffraction peaks could be independently fitted
by means of two Voigt functions of equal width
(figure~\ref{fig:peaks}). Due to the high background level and the
presence of Kikuchi lines, profiles associated with low titanium
coverage were difficult to decompose. In particular for coverage
below 2 nm, the existence of the $g_{\alpha 1}$ line could be
questioned. But at higher coverage of titanium the fits are
unambiguous. The fact that the in-plane atomic distances are
constant over a wide range of coverage (figure~\ref{fig:lat})
while the residue of the decomposition is very small
(figure~\ref{fig:peaks}) proves that the decomposition of the
diffraction line in two peaks is physically sound.

Figure~\ref{fig:lat} a) shows the evolution of the area of the
fitted peaks. In the 1-3 nm range, the $g_{\alpha 1}$ and
$g_{\omega 2}$ peak areas are about constant with $Ag_{\omega 2} >
Ag_{\alpha 1}$, while, above 3 nm, $g_{\alpha 1}$ increases and
$g_{\omega 2}$ progressively vanishes.  Due to the existence of
multiple diffraction effects in RHEED these intensities are only a
qualitative indications of the presence of the structures
corresponding to the observed peaks in the film. The relative
positions of the fitted peaks $g_{\alpha 1}$ and $g_{\omega 2}$
converted into the real space lattice constants $l_{\alpha}$ and
$l_{\omega}$ are presented in figure~\ref{fig:lat} b). Titanium
first grows with a lattice constant close to that of the substrate
$l_s$ and then undergoes a contraction of $\approx 4\%$ to a
lattice constant $l_{\omega}$ and finally it expands to the value
$l_{\alpha}$ close to the interatomic distance of bulk
Ti$[1\overline{1}00]$  which is $7.3\%$ larger than $l_s$, see
ref. \cite{crystals1}. The intermediate contraction of the film is
very surprising as the titanium lattice is already contracted on
the substrate, there should be no gain in energy by a further
contraction.

An analysis along the
$\makebox{Al}_2\makebox{O}_3[2\overline{11}0]$ zone axis also
reveals a change in lattice spacing of the titanium film even if
the diffraction patterns are more blurred than along the
$\makebox{Al}_2\makebox{O}_3[1\overline{1}00]$ axis
(figure~\ref{fig:axe2}). Around 2 nm,  $g_{\omega 1}$ streaks are
observed in a position which corresponds to a compression by $4\%$
with respect to the substrate. These streaks disappear while the
$g_{\alpha 2 }$ streaks become visible.

\subsection{Titanium as a buffer layer for silver growth}
\begin{figure}
\includegraphics[width=15cm]{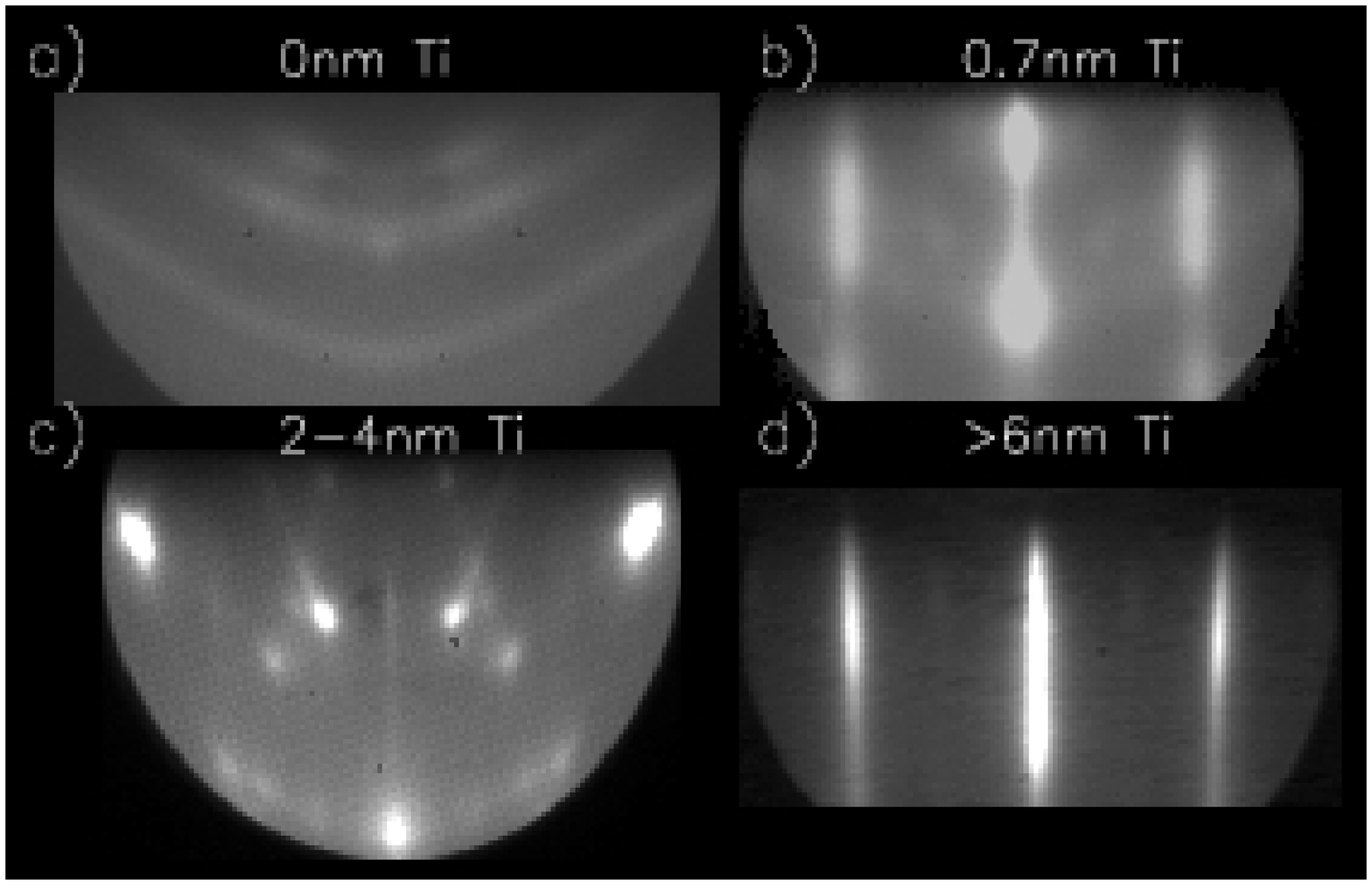}
\caption {The diffraction patterns of silver on various
Ti/$Al_2O_3$ films. Observed with the beam along
$Al_2O_3[1\overline{1}00]$.} \label{fig:silver}
\end{figure}

To elucidate the role of the thickness  of the  titanium buffer
effect, 2 nm thick silver films were grown with titanium buffers
of different thickness. Figure~\ref{fig:silver} shows typical
RHEED patterns of the silver films with the beam along the
$\makebox{Al}_2\makebox{O}_3[1\overline{1}00]$ axis. On the bare
$\makebox{Al}_2\makebox{O}_3\makebox{(0001)}$ substrate, between
300-600 K, silver grows without any epitaxial relation and
develops the ring structure typical for polycrystalline films
constituted of three dimensional domains (figure~\ref{fig:silver}
a). For a titanium layer of 0.5-1.5 nm, silver deposition leads to
streaks indicating a better spread of the film
(figure~\ref{fig:silver} b.). A full rotation in the plane shows
that the film has a sixfold symmetry corresponding to an epitaxy
of a Ag(111) plane. However, at $\pm 15^{\circ}$ around the
principal axes of the substrate, the diffraction pattern does not
change considerably and only decreases in intensity meaning that
the epitaxy is loose. On thicker titanium films (2-4 nm), when the
two streaks $g_{\omega 2}$ and $g_{\alpha 1}$ are present, the
silver layer exhibits a peculiar structure
(figure~\ref{fig:silver} c). The diffraction pattern is composed
of dots which suggest a bad wetting of the silver on substrate.
The pattern seen in figure~\ref{fig:silver} c) presents spots
which are characteristic of a long range order, but do not
correspond to any low index direction of silver. Finally, when
silver is deposited on a titanium film thicker than 4 nm and when
the $g_{\alpha 1}$ streak dominates, the diffraction streaks are
unperturbed by the silver deposition, which means that the silver
wets the substrate and grows epitaxially
(figure~\ref{fig:silver}d). Therefore, the behavior of the silver
film unambiguously shows that an important change occurs in the
titanium buffer between 4 and 5 nm.

\section{Discussion}

\subsection{Growth of titanium}
The data connected to the growth of the titanium film will first
be discussed. At high titanium coverage, the epitaxial
relationship is dominated by $\makebox{Ti}(0001)
\parallel \makebox{Al}_2\makebox{O}_3\makebox{(0001)} $,
$\makebox{Ti} [1 \overline{1}00]
\parallel \makebox{Al}_{2}\makebox{O}_{3}[2\overline{1}\overline{1}
0]$ and $\makebox{Ti}[10\overline{1}0]\parallel
\makebox{Al}_2\makebox{O}_3[1\overline{1}00]$ in agreement with
the epitaxy found by other groups for thicker films at much higher
temperature \cite{Ti2, Ti1}. But at lower coverage, extra
structures appear in the diffraction diagram in both zone axes in
the surface plane (figures ~\ref{fig:axe1} and ~\ref{fig:axe2}).
As the $g_{\omega 1}$ and the $g_{\omega 2}$ streaks show up in
the same coverage range and behave similarly, it is reasonable to
attach them to the same structure. The extra streaks can either be
due to multiple scattering between the principal streaks and an
overlying domaine structure or the existence of another epitaxial
relation between the substrate and the titanium film. Both
structures exhibit the same 6 fold symmetry as the substrate and
they are the only visible structures during a full rotation of the
sample. The dramatic change in the growth mode of the silver
overlayers is the indication of a significant difference between
the underlying structures of the titanium. This and the absence of
extra streaks around the zero order peak and the asymmetry of the
$g_{\omega 1}$ and the $g_{\omega 2}$ streaks make the existence
of a domaine structure unlikely.

If we assume that the observed pattern is the diffraction from a
crystal plane then the ratio $g_{\omega 2}/g_{\omega 2}= \sqrt{3}$
suggests a hexagonal structure with a lattice parameter $a \approx
\makebox{0.455}$ nm. The orientation of the hexagon is the same as
the hexagon formed by the outer Al atoms on the
$\makebox{Al}_2\makebox{O}_3(0001)$ surface. This plane does not
correspond to any hexagonal plane of the $\alpha$-Ti but it
compares with that of the (001) plane of the high pressure
hexagonal $\omega$ phase of titanium \cite{omegaphase}. Therefore,
the double peak observed in figure~\ref{fig:axe1} can be explained
by the  coexistence between two epitaxial hexagonal
configurations, one is $\alpha$ phase and the other is considered
to be the $\omega$ phase of titanium .

\begin{figure}
\includegraphics[width=16cm]{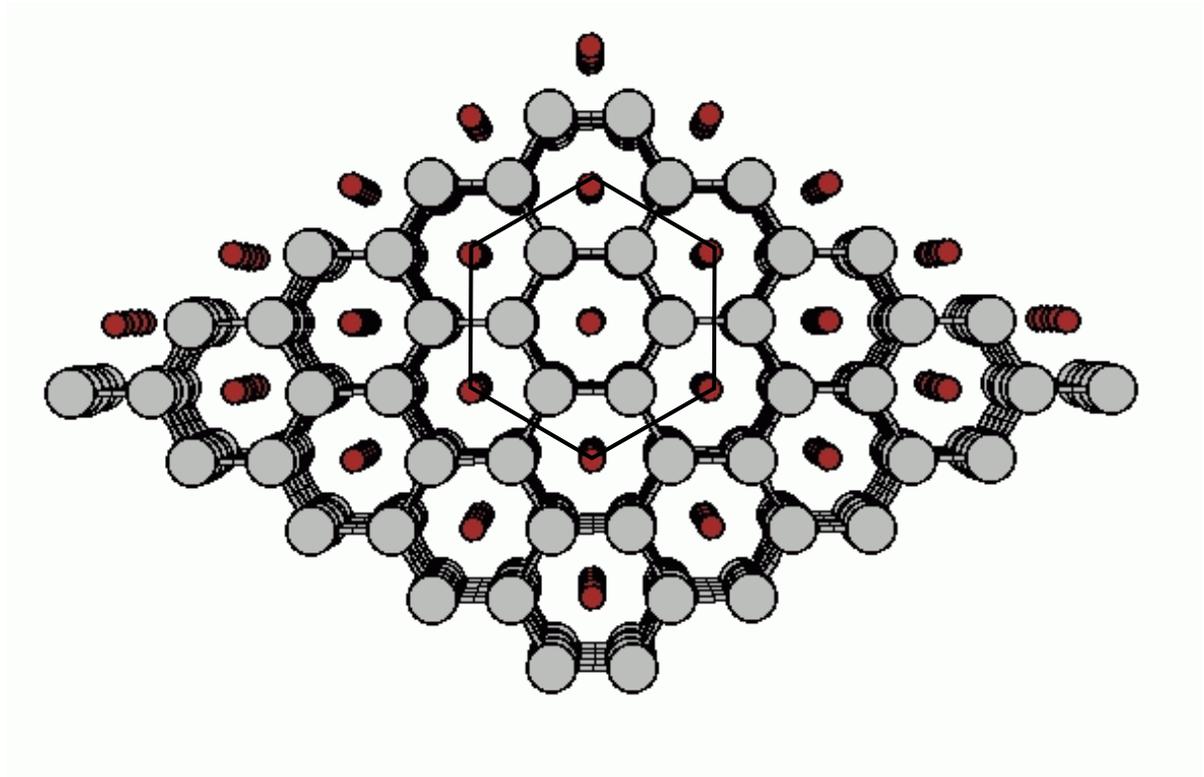}
\caption{Schematic top view of the $\omega$(0001) surface. The
structure can be seen as the successive stacking a triangular
lattice and a graphite like sheet. The indicated hexagon with an
edge of 0.455 nm corresponds to the underlying lattice.}
\label{fig:omega}
\end{figure}

The  $\omega$-Ti phase is a distorted bcc and one of the first
steps in the high pressure hcp to bcc phase transition exhibited
in the $IV$ metals \cite{omegahcp}. This phase is frequently
encountered in studies of the mechanical properties of titanium.
The different high pressure phases of titanium were recently
extensively studied both experimentally in a diamond anvil cell
\cite{omegaphase,omegaphase2} and theoretically
\cite{omegaphase3,omegaphase4}. Under hydrostatic pressure the
$\omega$ phase coexists with the $\alpha$ phase from around 2 GPa
to 9 GPa where the $\omega$ phase dominates. The phase is stable
up to 116 GPa. Once the $\omega$ phase is obtained it can be kept
at the ambient pressure although it is metastable. The
$\omega$-Ti(001) plane has been schematized (
figure~\ref{fig:omega}) using the lattice parameters given in Ref.
\cite{omegaphase3} and the structure description in Ref.
\cite{NRLsite}. The $\omega$ phase can be seen as the successive
stacking of a hexagonal lattice and a graphite like sheet, the
hexagon being the origin of the periodicity of the lattice. The
parameter of this hexagon a=0.4598nm \cite{NRLsite} is indeed
close to what was observed herein.

A few substrate-induced phase transitions have been observed in
metal on metal epitaxial systems. One of most studied exemples is
the fcc - bcc martensitic transition of iron on metallic
substrates like Cu and $Cu_3Au(001)$. A very elaborated exemple is
the STM and RHEED study of the Bains transition of iron on
$Cu_3Au(001)$, \cite{Fe0,Fe1,Fe2}. As in the present case, the
authors obtained a critical thickness for the fcc phase between 1
and 10 ML depending on the substrate preparation and the films
exhibited a high amount of in-plane disorder.

\begin{figure}
\includegraphics[width=15cm]{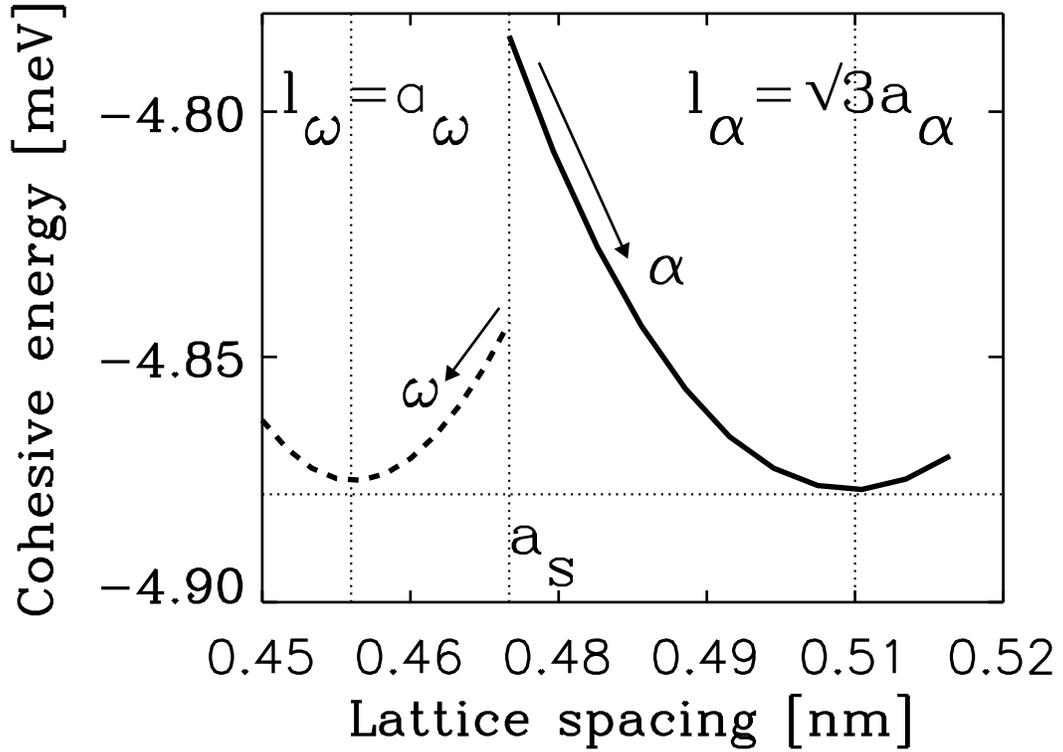}
\caption{ Calculated total energy  per atom of $\alpha$-Ti (full
line) (resp. $\omega$-Ti (dashed line)), for lattice parameters
ranging from the one giving the perfect lattice matching with the
substrate ({\sl i.e.} $a_{\alpha}=0.475/\sqrt{3}$ nm for
$\alpha$-Ti and $a_{\omega}=0.459$\AA for $\omega$-Ti), up (down)
to the equilibrium lattice constant. Note that the total energy of
$\alpha$-Ti is represented as a function of $l_{\alpha}=\sqrt{3}
a_{\alpha}$ whereas the total energy of $\omega$-Ti is represented
as a function of $l_{\omega}= a_{\omega}$.}
\label{fig:energ_hcp_omega}
\end{figure}

To evaluate the energetic balance between the $\alpha$ and
$\omega$ phases under various compressive and extensive conditions
a serie of total energy calculations was undertaken. The very
small energy difference between the $\alpha$ and $\omega$ phase
makes such calculations difficult. As a matter of fact, this phase
transition is not properly described by first-principles methods
by which the equilibrium $\omega$-Ti phase is found to be slightly
lower in energy (about 7 meV) than $\alpha$-Ti \cite{omegaphase3}.
Therefore, a total energy calculations was performed using the set
of tight-binding parameters  recently developed by
Papaconstantopoulos and co-workers \cite{omegaphase3,code} to
reproduce as accurately as possible the $\alpha$ to $\omega$ phase
transition. The calculated equilibrium lattice constants of the
$\alpha$-Ti and $\omega$-Ti phases are $a_{\alpha}=0.294$ nm ({\sl
i.e}, $l_{\alpha}=\sqrt{3}a_{\alpha}=0.51$ nm) and
$a_{\omega}=0.459$ nm respectively, showing a lattice mismatch
with respect to the substrate of $+7.4\%$ and $-3.4\%$.  The total
energy per atom of $\alpha$-Ti (resp. $\omega$-Ti) has been
calculated for hexagonal lattice parameters ranging from perfect
matching with the substrate, up (resp. down) to the equilibrium
lattice constants (figure \ref{fig:energ_hcp_omega}). Each point
of the curve is obtained by fixing the hexagonal lattice parameter
$a$ and minimizing the total energy with respect to  the parameter
$c$. It clearly appears that, at the  lattice constant giving the
perfect matching with the substrate, $\omega$-Ti is favored with
respect to $\alpha$-Ti, whereas it is the reverse at the
equilibrium lattice constant. The small energy difference between
the two states makes the coexistence between the two phases over a
range of deposited thicknesses  very plausible. The presence of
defects on the substrate surface can also modify the energetic
balance between the two phases. The RHEED data do not give any
indication about the arrangement of the two phases which are
assumed to coexists within the titanium layer nor does it  give
insight about the burried part of for the titanium. However, data
show clearly that above a certain thickness the surface of the
titanium film is exclusively made of the $\alpha$- Ti with the
$\makebox{Ti}(0001)
\parallel \makebox{Al}_2\makebox{O}_3\makebox{(0001)} $,
$\makebox{Ti} [1 \overline{1}00]
\parallel \makebox{Al}_{2}\makebox{O}_{3}[2\overline{1}\overline{1}
0]$ and $\makebox{Ti}[10\overline{1}0]\parallel
\makebox{Al}_2\makebox{O}_3[1\overline{1}00]$ epitaxy.

\subsection{Influence of the cristallography of titanium on the wetting of silver}
Concerning the growth of silver on titanium the strength of the
metallic bond should clearly favor the good spreading of silver on
titanium. Furthermore the lattice spacing the of two most compact
hexagonal planes of titanium and silver are the Ti(0001) with $a =
0.295$ nm and the Ag(111) with $a = 0.2889$ nm \cite{crystals1}.
Therefore, it can be anticipated that the Ag/Ti epitaxy is favored
over the Ag/$\makebox{Al}_2\makebox{O}_3$ epitaxy because it
corresponds to a lower mismatch (-2\% versus 5.34\%). Thus, both
the surface energetic and the lattice constants seems to be in
favor of a better wetting of silver on titanium on
$\makebox{Al}_2\makebox{O}_3$ at least when bulk values are
considered.

The data presented in figure~\ref{fig:silver} present a more
complex picture and suggest that a part of the buffer effect of
titanium on silver is linked to the structure and the stress of
the interface. On the bare alumina substrate silver grows as three
dimensional polycrystals. Silver does not wet the alumina which
transfers little order to the metal overlayer. However, the
presence of very thin titanium films improves the wetting of
alumina by silver ($\approx$ 1 nm, see Ref. \cite{SGR}, the silver
film showing a texture around Ag(111) orientation meaning that the
titanium film transfers a part of the substrate order. Taking into
account the very diffuse diffraction patterns for thin Ti layers
in this region it is possible that the silver film only replicates
the underlying Ti structure.  At higher coverage, for titanium
layers where $g_{\omega 2}$ dominates and the $g_{\alpha 1}$ just
starts to become visible, the wetting of the substrate by silver
becomes poor, figure~\ref{fig:silver}c. Finally, when the titanium
film is thicker and has a nearly relaxed lattice parameter, silver
grows epitaxially in a laminar way. The $\mbox{Ag}(111) \|
\mbox{Ti}(0001), \mbox{Ag}[110] \| \mbox{Ti}[2\overline{11}0]$
epitaxy corresponds to expectation and it is similar to that found
for the Cu/Ti/$\mbox{Al}_2\mbox{O}_3$ system \cite{Ti2}. Such a
structure is consistent with the observation of a pure $\alpha$ Ti
phase by electron diffraction. The bad wetting of silver on the
intermediate titanium coverage is an evidence that a titanium
structure drastically different from $\alpha$ Ti is present.
Indeed, a like explanation is the presence of the $\omega$ Ti at
the surface of the titanium film. The change in the interfacial
energy is due the large mismatch (9.9\%) between $\omega$ Ti(0001)
and Ag(111) and prevents the spreading of silver.

The present data show that  the silver layers  exhibit a
remarkable amount of different textures. Laminar growth is only
obtained when the lattice constant of the titanium film is very
close to the silver bulk value. Therefore, the buffer effect of
titanium for noble metals is not only the result of a change in
surface and interface energies. It is also very dependent on the
crystallography of the buffer and therefore on the elastic
contributions to the interfacial energy.

\section{Conclusion}
The room temperature growth of thin titanium films on $\alpha -
\makebox{Al}_2\makebox{O}_3(0001)$ has been shown be more complex
than hereto reported. For films above 4 nm the $\alpha$-Ti phase
is obtained in an epitaxial relation  already observed and
described in literature for the high temperature growth of thick
titanium films on sapphire. For thin films an additional structure
is present. This structure is assigned to the (0001) plane of the
$\omega$ phase where the hexagon is aligned with the one of an
Al-terminated $\alpha - \makebox{Al}_2\makebox{O}_3(0001)$. The
two phases coexist  at the beginning of the growth of the titanium
film in the range of coverage ($\approx$ 1-4 nm). The
substrate-induced existence of an $\omega$ phase is plausible when
taking simple lattice considerations into account, but the details
behind the growth and stabilisation of this phase still need to be
explained.

The structure of the titanium film has a dramatic influence on the
so-called buffer effect, which has been tested herein by
depositing silver on a $\alpha -
\makebox{Al}_2\makebox{O}_3\makebox{(0001)}$ surface pre-covered
by titanium at various coverage. Indeed, the titanium film
structure and strain was proved to strongly affect, not only the
structure, but also the wetting of the silver film. The silver
film, which clearly grows in a 3D manner when the film structure
contains the structure we assign to the $\omega$-Ti phase, spreads
out on the $\alpha$-Ti phase. Therefore, the growth mode and
spreading of noble metals deposited on titanium-covered alumina
arises not only from thermodynamics but from a subtle interplay
between the structure and the surface energy of the titanium film.

\section{Acknowledgements}
We wish to thank the group of Dr. Claude Fermon at the CEA-France
for the use of their RHEED/UHV facility. Furthermore Dr. A. Marty
gave helpful comments during the redaction of this paper. Dr. C.
Barreteau wish to thank the Drs. D.A. Papaconstantopoulos and N.
Bernstein at the Naval Research Center for the access to their
tight binding code and the stay at NRL founded by an ONR grant.
The group of Dr. Patrice Lehuede, Saint Gobain Recherche, helped
with the electron microprobe measurements.

\cleardoublepage

%

%

%

\end{document}